# Field emission from single and few-layer graphene flakes

S. Santandrea[1], F. Giubileo[1], V. Grossi[2], S. Santucci[2], M. Passacantando[2], T. Schroeder[3], G. Lupina[3], and A. Di Bartolomeo[1]

[1] Dipartimento di Fisica, Università degli Studi di Salerno and CNR-SPIN Salerno, and Centro NANO_MATES, via Ponte don Melillo, 84084 Fisciano (SA), Italy

[2] Dipartimento di Fisica, Università degli Studi dell'Aquila, via Vetoio, 67010 Coppito, L'Aquila, Italy

[3] IHP, Im Technologiepark 25, 15236 Frankfurt (Oder), Germany

We report the observation and characterization of field emission current from individual single- and few-layer graphene flakes laid on a flat $SiO_2$/Si substrate. Measurements were performed in a scanning electron microscope chamber equipped with nanoprobes, used as electrodes to realize local measurements of the field emission current. We achieved field emission currents up to 1 µA from the flat part of graphene flakes at applied fields of few hundred V/µm. We found that emission process is stable over a period of several hours and that it is well described by a Fowler-Nordheim model for currents over 5 orders of magnitude.

Field-emission (FE) is a quantum tunneling phenomenon in which electrons pass from an emitting material (cathode) to an anode through a vacuum barrier by effect of high electric field. For a given material, cathodes with higher aspect ratios and sharper edges produce higher FE currents. Nanostructures, with very high aspect ratio, are considered promising for commercial applications as in flat-panel displays, vacuum electronics, microwave power tubes, electron sources, etc.

Graphene, consisting of a single- or few-graphite layers, at moment a very popular nanostructure in material science[1-3], shares many similar or even superior properties as carbon nanotubes (CNTs)[4-6], and as CNTs can have high potentiality for FE applications[7-9]. Due to the very high aspect ratio (thickness to lateral size), a dramatically enhanced local electric field is expected at the edges of graphene flakes. So far, most of the work has been performed on graphene films[10-15] or on graphene/polymer composites[16] reporting FE from graphene edges or pleats at low fields. To date, no observation of FE current from the inner, flat part of graphene flakes has been reported[17].

From a theoretical point of view, FE experiments can be analyzed in terms of Fowler–Nordheim (FN) theory[18], the most-commonly used model for FE from a metallic or semiconductor surface under a strong applied field, that has been also widely used to investigate the FE from CNTs[7,19-23]. According to such model, the current density (J) produced by a given electric field (E) is described by the equation:

$$J = \frac{A\beta^2 E^2}{\phi}\exp\left(-\frac{B\phi^{3/2}}{\beta E}\right), \quad I = S \cdot J, \quad E = \frac{V}{d} \quad (1)$$

where A and B are constants, (A = $1.54 \cdot 10^{-6}$ A eV V$^{-2}$, B = $6.83 \cdot 10^7$ eV$^{-3/2}$ V cm$^{-1}$), β is the field enhancement factor at the emitting surface, *d* is the distance between the sample and the anode, $\phi$ is the work function of the emitting materials, *I* is the emission current, *S* the emitting area and V the applied voltage. According to (1), the FN behavior predicts a straight line plotting $\ln(I/V^2)$ versus $1/V$. However, to justify recent observation of non linear FN plots, it has been proposed[17] that FE processes can undergo a low-to-high field regime transition that could provoke an up-bending of the FN plot, suggesting the necessity of introducing a more general relation $\left(\ln(I/E^\alpha) \approx 1/E^\beta\right)$ to account for FE from graphene.

In this paper, we report, as far as we know, the first direct observation of FE current from the inner, flat part of single- and few- layer graphene flakes. Taking advantage of a special setup, consisting of two nanomanipulator in a SEM chamber connected to an external source measurement unit, we investigate FE currents by applying electric fields up to 2 kV/µm. We show that a high and stable FE current (up to 1 µA) well described by the usual F-N model over several order of magnitude, can be achieved with a turn-on field of ~600 V/µm.

Single- and few-layer graphene sheets were prepared by mechanical exfoliation of highly ordered pyrolytic graphite and transferred by scotch-tape method on a layer of $SiO_2$ thermally grown on top of highly p-doped Si substrate. The thickness of the $SiO_2$ layer was carefully chosen to ~300 nm to make the single and few-layer graphene clearly visible and distinguishable under optical microscope, which was used for a first localization of the flakes in the reference frame of a regularly 100 µm-spaced grid of Au crosses deposited on the substrate by conventional lithography (Figure 1a). Raman spectroscopy was further exploited to give confirmation of the single- or few-layer graphene (Figure 1b).

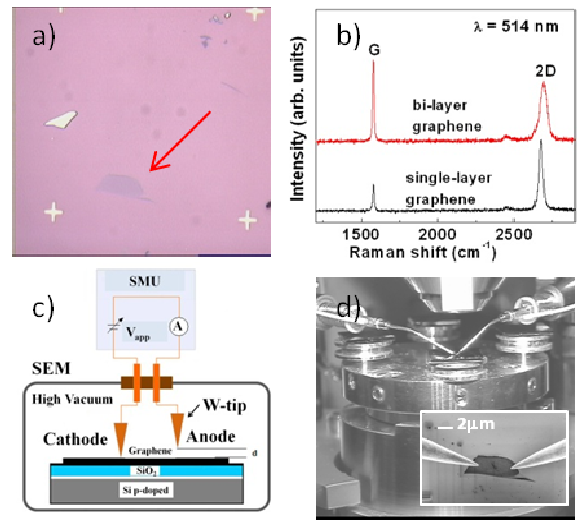

**Figure 1.** a) Optical image of a graphene flake at 50x magnification; b) Typical Raman spectra measured for single- and bi- layer graphene; c) Schematic view of the experimental setup using a SMU interconnected to nanomanipulators in SEM chamber. d) Photograph of the inner SEM chamber, and Inset: SEM image of graphene flake of fig. 1a contacted by the nanomanipulators.

The FE measurements were performed inside a SEM chamber equipped with two nanomanipulators, ending with tungsten tips with curvature radius <100nm. The SEM chamber was kept to a pressure of $10^{-8}$ Torr and at room temperature. The tips, one in contact to graphene and the second at a varying distance from it, were used as the cathode and the anode, respectively. Vacuum feed-throughs connected the two nanoprobes to a source-measurement unit (SMU).

In figure 1c we show a scheme of the measurement setup, while a real image of the inner part of SEM chamber is reported in figure 1d, with the tips pointing to the sample.

As measurement procedure, we first imaged a graphene flake in the SEM profiting from the previous localization with respect Au markers under optical microscope. Then, we gently contacted the flake with both nanoprobes, paying attention at not provoking breakage of the graphene lattice, and we measured its I-V characteristics. The series resistance of the measurement loop obtained with the two tips shorted was 3Ω. When the tips were approaching a graphene flake, the measured resistance was observed to decrease from MΩ till kΩ range. We observed that successive voltage sweeps acted as electrical conditioning, appreciably improving the quality of the nanoprobe-graphene contacts.

Figure 2a shows an example of I–V curves when both the probes were in contact with a graphene flake. Sweep 1 was reproducibly measured by limiting the maximum current to 50 µA; as soon as we forced higher current (200 µA) through the contacts, the circuit resistance was sensibly reduced (sweep 2); the third sweep is the result after electrical stress with 300 µA current. Higher current values were avoided in order to prevent graphene flake breaking. Figure 2a demonstrates that increasing the maximum current improves the contact, thus lowering the resistance and making the current more stable. The I-V characteristics of figure 2a are only slightly nonlinear indicating that approximately Ohmic contacts are established.

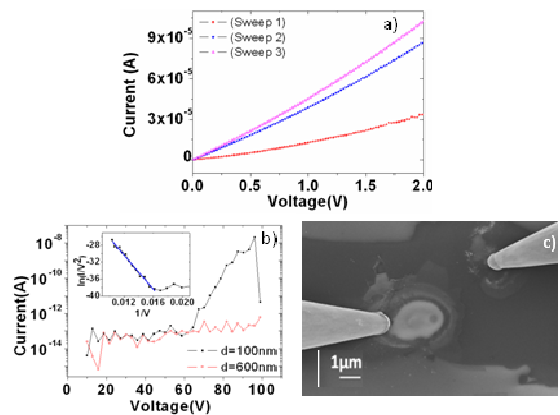

**Figure 2.** a) *I-V* characteristics obtained after allowing different values of maximum current, i.e. different electrical stress; b) V curve for FE measurements at two different graphene-tip distances (100 nm, 600 nm): inset shows the FN plot; c) Electrical breakdown resulting in melted tips and destroyed graphene as effect of high current.

FE measurements have been then performed by pulling one electrode (the anode) away from the graphene flake and recording I-V characteristics for voltages ranging up to 200 V. Voltage sweeps were performed while monitoring the current at different distances of the anode from the flake (keeping other probe in contact). In order to avoid vacuum breakdown or damage of the graphene flake, we kept current limit below 10 µA.

Figure 2b shows the current-voltage characteristics for a few-layer graphene flake at two different anode-flake distances, d~100 nm and d~600 nm. For the larger distance, we did not observe any FE current flowing between the electrodes. For reduced distance (d~100 nm) a current at V > 60 V exponentially increases for more than 5 orders of that corresponds to a linear FN plot confirming its FE nature. Importantly, this confirms that FE current from a flat graphene layer can be described by FN model, no modified model being necessary to account for experimental observations. We repeated several measurements always in good agreement with equation (1) .

The turn on field, here defined as the field needed to extract a current of 10 pA, is $E_{ON} \approx 600 \frac{V}{\mu m}$. This high field is not surprising since the electrons are emitted from a flat surface. We stress that, to the best of our knowledge, this is the first time that FE is observed from the flat, inner part of a graphene flake.

Rising the voltage, and therefore the current in the circuit, may result in a catastrophic event characterized by the melting of the tungsten tips and the local breakage of the flake. Indeed, the high current provokes high power dissipation at the graphene-tip contact (due to high resistance of the constriction), and the local heating can melt the tungsten tip and locally destroy the graphene flake (figure 2c). This phenomenon is electrically observed as a sudden drop of the FE current (last point in the I-V curve of figure 2b) resulting in an open circuit.

More importantly, we succeeded to perform FE measurements from single layer graphene. In figure 3a we show typical FE measurement, where the anode, initially in contact, was moved away from the flake at steps of ~10 nm. SEM image of the flake that we previously characterized by optical and Raman investigation is reported in the inset of figure 3b. Sweep-1 and sweep-2 refer to a contact condition with a high graphene/tip contact resistance; sweep-3 shows a change of behavior with FE current starting around 55 V and the graphene-tip contact soon re-established few volts later as effect of electrostatic attraction; moving the tip further away (sweep-4) results in the observation of FE current above 100V and for the entire voltage range up to 175V. We point out that FE current in sweep-4 spans 6 orders of magnitude, and it follows the standard FN model with clear linear dependence, as shown in figure 3b.

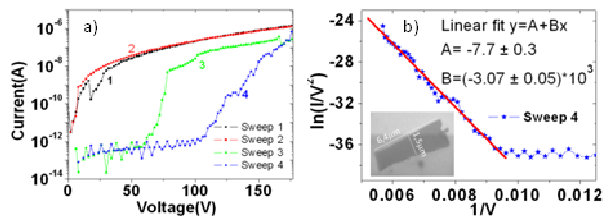

**Figure 3.** a) SEM image of a single layer graphene; b) I-V curves measured for increasing graphene-anode distance; c) linear FN plot for FE characteristic (sweep 4).

Repeated voltage sweeps have been shown to stabilize the FE current; in particular for CNT emitters it has been shown that the electrical stabilization results also in an increase of the turn-on voltage as consequence of desorption of adsorbates and, in case of CNT films, of burning of the most protruding CNTs with respect the whole forest of tubes[24]. For graphene, we found that electrical conditioning still stabilizes the FE current which achieves a stable common value at high fields (figure 4a); however, a different trend is observed for single layer graphene with respect to CNTs since successive sweeps reduce the turn-on field. This reverse trend can be understood considering that repeated sweeps improve the cathode-graphene contact thus changing the partition of the applied voltage among the cathode-graphene and the anode-graphene resistances. The gradually reduced cathode-graphene contact resistance makes the applied voltage mainly drop at the FE (anode-graphene) junction, where a much higher effective applied field promptly triggers FE current.

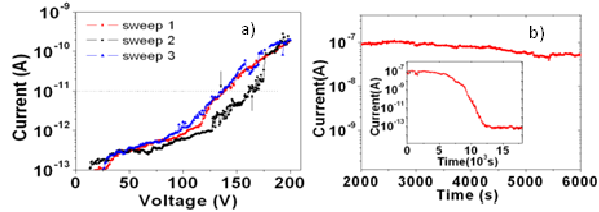

**Figure 4.** a) FE current versus bias voltage for successive sweeps at a graphene-anode nano probe distance ~1μm. Arrows indicate the turn on level; b) FE current vs time. Applied voltage 100 V. Insert shows the FE over a larger time scale.

Finally, we tested the emission stability from the flat inner part of single-layer graphene flake, due its for practical applications. Experimentally, we applied a constant bias and monitored the current over periods of several hours. Figure 4b shows the result obtained by measuring the FE current every 10 s under continuous bias of 100 V over a time of about 2 h. A good stability was obtained. The insert of figure 4b shows that for longer times thermal drift caused a gradual increase of the graphene-anode distance, leading to a current decrease till its complete suppression.

We have reported the first observation of field emission current from the inner, flat part of single- and few-layer graphene. The current drawn can reach values up to 1μA and is triggered with a turn-on field of ~ 600 V/μm. We have shown that FE current follows the F-N model over 5 orders of magnitude. Our setup allowed monitoring of FE only for few hours, over which a stable process was confirmed.